\documentclass{aa}  
\usepackage{natbib}
\usepackage{dcolumn}
\newcolumntype{d}[1]{D{.}{.}{#1}}
\usepackage{multirow}
\usepackage{listings}
\usepackage{array}
\usepackage[normalem]{ulem}
\newcommand{\mydistance}{+0.5ex}
\usepackage{amssymb}
\usepackage{enumitem} 
\usepackage{amsmath}
\usepackage{float}
\usepackage{epsfig}
\usepackage{placeins}

\usepackage{color}
\usepackage{graphicx}
\usepackage{txfonts}

\begin{document}

   \title{Accurate reference spectra of HD in H$_2$/He bath\\ for planetary applications}

   \author{H.~J\'{o}\'{z}wiak
          \inst{1}, N.~Stolarczyk\inst{1}, K.~Stankiewicz\inst{1}, M.~Zaborowski\inst{1}, D.~Lisak\inst{1}, S.~Wójtewicz\inst{1},  P.~Jankowski\inst{2}, K.~Patkowski\inst{3}, K.~Szalewicz\inst{4}, F.~Thibault\inst{5}, I.E.~Gordon\inst{6}, P.~Wcisło\inst{1}
          }

   \institute{Institute of Physics, Faculty of Physics, Astronomy and Informatics, Nicolaus Copernicus University in Toru\'{n}, Grudziadzka 5, 87-100 Toru\'{n}, Poland
         \and
         Faculty of Chemistry, Nicolaus Copernicus University in Toru\'{n}, Gagarina 7, 87-100 Toru\'{n}, Poland
         \and 
        Department of Chemistry and Biochemistry, Auburn University, Auburn, Alabama 36849, USA
         \and
         Department of Physics and Astronomy, University of Delaware, Newark, Delaware 19716, USA
         \and
             Univ Rennes, IPR (Institut de Physique de Rennes) – UMR 6251, F-35000 Rennes, France 
        \and
        Harvard-Smithsonian Center for Astrophysics, Atomic and Molecular Physics Division, Cambridge 02138, USA
             }
 
  \abstract
   {The hydrogen deuteride (HD) molecule is an important deuterium tracer in astrophysical studies. The atmospheres of gas giants are dominated by molecular hydrogen, and simultaneous observation of H$_2$ and HD lines provides reliable information on the D/H ratios on these planets. The reference spectroscopic parameters play a crucial role in such studies. Under thermodynamic conditions encountered in these atmospheres, the spectroscopic studies of HD require not only the knowledge of line intensities and positions but also accurate reference data on pressure-induced line shapes and shifts.
   }
   {Our aim is to provide accurate collision-induced line-shape parameters for HD lines that cover any thermodynamic conditions relevant to the atmospheres of giant planets, i.e., any relevant temperature, pressure, and perturbing gas (the H$_2$/He mixture) composition.}
   {We perform quantum-scattering calculations on our new highly accurate \textit{ab initio} potential energy surface, and we use scattering S-matrices obtained this way to determine the collision-induced line-shape parameters. We use the cavity ring-down spectroscopy for validation of our theoretical methodology.}
   {We report accurate collision-induced line-shape parameters for the pure rotational R(0), R(1), and R(2) lines, the most relevant HD lines for the investigations of atmospheres of the giant planets. Besides the basic Voigt-profile collisional parameters (i.e., the broadening and shift parameters), we also report their speed dependences and the complex Dicke parameter, which can influence the effective width and height of the HD lines up to almost a factor of 2 for giant planet conditions.
   The sub-percent-level accuracy, reached in this work, considerably improves the previously available data. All the reported parameters (and their temperature dependences) are consistent with the HITRAN database format, hence allowing for the use of HAPI (HITRAN Application Programming Interface) for generating the beyond-Voigt spectra of HD.}
   {}


   \authorrunning{H.~J\'{o}\'{z}wiak et al.}
   \maketitle
%
\section{Introduction}
\label{sec:Intro}
Hydrogen deuteride (HD), the second most abundant isotopologue of molecular hydrogen, is an important tracer of deuterium in the Universe. The small and constant primordial fraction of deuterium to hydrogen, D/H ($(2.8 \pm 0.2) \times 10^{-5}$,~\citet{Pettini_2008}), is one of the key arguments supporting the Big Bang theory. Measurements of the D/H ratio in the Solar System provide information about planetary formation and evolution.  The standard ratio on Earth is considered to be Vienna Standard Mean Ocean Water (VSMOW; D/H = 1.5576 $\times$ 10$^{-4}$, \citet{Araguas-Araguas1998}). With that \citet{Donahue_1982} reported the D/H ratio in Venusian atmosphere to be $(1.6\pm0.2) \times10^{-2}$ (i.e. two orders of magnitude higher than VSMOW). This higher ratio is attributed to the evaporation of oceans and subsequent photodissociation of H$_{2}$O in the upper parts of the atmosphere~\citep{Donahue_1983}. Comparison of the D/H ratio on Earth and Mars with the values determined for various comets indicate the role of the latter in the volatile accretion on these two planets~\citep{Drake_2002, Hartogh_2011}. The two largest gas giants, Jupiter and Saturn, are incapable of nuclear fusion, thus planetary models predict that the D/H ratio in their atmospheres should be close to the primordial value for the Solar System~\citep{Lellouch_2001} or slightly larger, due to the accretion of deuterium-rich icy grains and planetesimals~\citep{Guillot_1999}. The abundance of deuterium is larger by a factor of 2.5 in the atmosphere of Uranus and Neptune~\citep{Feuchtgruber_2013}, owing to their ice-rich interior~\citep{Guillot_1999}. Additionally, the determination of the D/H ratio in comets and moons provides information about the formation of ices in the early Solar System~\citep{Hersant_2001,Gautier_2005,Horner_2006}. 

The D/H ratio in planets, moons, or comets can be derived from either \textit{in situ} mass spectrometry \citep{Eberhardt_1995, Mahaffy_1998, Niemann_2005, Altwegg_2014} or spectroscopic observations of various molecules and their deuterated isotopologues in millimeter and infrared range~\citep{Bockele_Morvan_1998, Meier_1998, Crovisier_2004, Fletcher_2009, Pierel_2017, Krasnopolsky2013, Blake_2021}. Interestingly, as pointed out by \citet{Krasnopolsky2013}, the ratios determined from spectra of different molecules can differ substantially. Moreover, even on Earth, the HDO/H$_2$O ratio in the atmosphere can vary substantially across the globe (\citet{Araguas-Araguas1998}). In this context, the observation of isotopologues of molecular hydrogen can be a reliable benchmark for determining D/H ratio. The most accurate values of the D/H ratio in gas giants stem from the analysis of the pure rotational R(0), R(1), R(2), and R(3) lines of HD and S(0) and S(1) lines in H$_{2}$. \citet{Lellouch_2001} determined the D/H ratio in the Jovian atmosphere using the Short Wavelength Spectrometer (SWS) on board the Infrared Space Observatory to be ($2.25\pm0.35) \times 10^{-5}$. \citet{Pierel_2017} analyzed the far-infrared spectra of Saturn’s atmosphere gathered by Cassini’s Composite Infrared Spectrometer. Interestingly, Pierel’s result suggests that the D/H ratio on Saturn is lower than that of Jupiter, which contradicts the predictions based on interior models \citep{Guillot_1999, Owen_2006}, and points to an unknown mechanism of fractionating deuterium in Saturn’s atmosphere. The D/H ratio in the atmospheres of Uranus and Neptune was determined from measurements of the pure rotational R(0), R(1), and R(2) lines in HD using the PACS spectrometer on board Herschel space observatory \citep{Feuchtgruber_2013}. The analysis revealed similar values for these two giants ($(4.4\pm0.4)\times10^{-5}$ and $(4.1\pm0.4)\times10^{-5}$, respectively), confirming the expected deuterium enrichment with respect to the protosolar value.

Atmospheric models of the Solar System’s gas giants, from which the relative abundance of HD with respect to H$_{2}$ (and consequently, the D/H ratio) is retrieved, require knowledge about the line parameters of HD and H$_{2}$. Temperature profiles, considered in the models, cover seven orders of magnitude of pressure (1~$\mu$bar–10~bar, see for instance~\citet{Pierel_2017}). Thus, apart from line position and intensity, knowledge about collisional effects that perturb the shape of observed lines is crucial for the accurate determination of the relative HD abundance. Furthermore, the incorporation of non-Voigt line-shape effects (such as Dicke narrowing and speed-dependent effects) allows for reducing the systematic errors in atmospheric models, as shown for Jupiter~\citep{Hayden_Smith_1989}, and Uranus and Neptune~\citep{Baines_1995}. Indeed, although the spectral lines of most molecules observed in planetary atmospheres are not sensitive to the non-Voigt effects considering the resolving power of available telescopes, the lines of molecular hydrogen and its isotopologues are~\citep{Hayden_Smith_1989, Baines_1995}. 

In this article, we report accurate collision-induced line-shape parameters for the rotational transitions R(0), R(1), and R(2) within the ground vibrational state. These HD lines are most frequently used for studies of the atmospheres of the giant planets. The results cover any thermodynamic conditions relevant to the atmospheres of giant planets in the solar system, i.e., any relevant temperature, pressure, and the H$_2$/He perturbing gas composition (the HD-H$_2$ data are provided in this work, while the HD-He data are taken from \citet{Stankiewicz_2020, Stankiewicz_2021}). Besides the basic Voigt-profile collisional parameters (i.e., the broadening and shift parameters), we also report their speed dependences and the complex Dicke parameter, which, as we will show, can influence the effective width and height of the HD lines up to almost a factor of 2 for giant planet conditions. For the R(0) line, the non-Voigt regime coincides with the maximum of the monochromatic contribution function (see Fig.~1 in \cite{Feuchtgruber_2013}), which has a direct influence on the abundance of HD inferred from observations.

We perform quantum-scattering calculations on our new highly accurate \textit{ab initio} 6D potential energy surface (PES), and we use the scattering S-matrices obtained in this way to determine the collision-induced line-shape parameters. We use the cavity ring-down spectroscopy for accurate validation of our theoretical methodology, demonstrating sub-percent-level accuracy that considerably surpasses the accuracy of any previous theoretical~\citep{Schaefer_1992} and experimental studies of line-shape parameters of pure rotational lines in HD~\citep{Ulivi_1989, Lu_1993,Sung_2022}, and offers valuable input for the HITRAN database~\citep{Hitran_2020}. This work presents significant methodological and computational progress; calculations at this level of theory and accurate experimental validation have been formerly performed for a molecule-atom system (3D potential energy surface) \citep{Slowinski_2020, Slowinski_2022}, while in this work, we extend it to a molecule-molecule system (6D potential energy surface). All the reported parameters (and their temperature dependences) are consistent with the HITRAN database format, hence allowing for the use of HAPI (HITRAN Application Programming Interface)~\citep{Kochanov2016} for generating the beyond-Voigt spectra of HD for any H$_2$/He perturbing gas composition and thermodynamic conditions.

\section{\textit{Ab initio} calculations of the line-shape parameters}
\label{sec:Theory}

In recent years, the methodology for accurate \textit{ab initio} calculations of the line-shape parameters (including the beyond-Voigt parameters) was developed and experimentally tested for He-perturbed H$_2$ and HD rovibrational lines, starting from accurate \textit{ab initio} H$_2$-He potential energy surface calculations \citep{Bakr_2013,Thibault_2017}, through the state-of-the-art quantum scattering calculations and line-shape parameter determination \citep{Thibault_2017,Jozwiak_2018}, up to accurate experimental validation \citep{Slowinski_2020, Slowinski_2022} and using the results for populating the HITRAN database \citep{Wcislo_2021,Stankiewicz_2021}.

In this work, we extend the entire methodology to a much more complex system of a diatomic molecule colliding with another diatomic molecule. First, we calculated an accurate 6D H$_2$-H$_2$ PES. Second, we performed state-of-the-art quantum-scattering calculations. Third, we calculated the full set of the six line-shape parameters in a wide temperature range. Finally, we report the results in a format consistent with the HITRAN database.

The 6D PES was obtained using the supermolecular approach based on the
level of theory similar to that used to calculate the 4D H$_{2}$-H$_{2}$ surface ~\citep{Patkowski:08}. The crucial contributions involve interaction energy calculated at the Hartree-Fock (HF) level, the correlation contribution to the interaction energy calculated using the coupled-cluster method with up to perturbative triple excitations, CCSD(T) (with the results extrapolated to the complete basis set (CBS) limit~\citep{Halkier:99}), electron correlation effects beyond CCSD(T) up to full configuration interaction (FCI), and the diagonal Born-Oppen\-hei\-mer correction (DBOC)~\citep{Handy:86}. The details regarding the basis sets used in calculations of each contribution and the analytical fit to the interaction energies are given in Appendix~\ref{appendix:PES}. The PES is expected to be valid for intramolecular distances $r_{i} \in [0.85, 2.25]\,a_{0}$.

For the purpose of performing quantum scattering calculations, the 6D PES is expanded over a set of appropriate angular functions and the resulting 3D numerical function in radial coordinates is then expanded in terms of rovibrational wave functions of isolated molecules, see Appendix~\ref{app:quantum_scattering} for details. The close-coupling equations are solved in the body-fixed frame using a renormalized Numerov's algorithm, for the total number of 3\,014 energies ($E_{T} = E_{\rm{kin}}+E_{v_{1}j_{1}}+E_{v_{2}j_{2}}$, where $E_{\rm{kin}}$ is the relative kinetic energy of the colliding pair, and $E_{v_{1}j_{1}}$ and $E_{v_{2}j_{2}}$ are the rovibrational energies of the two molecules at large separations) in a range from  10$^{-3}$~cm$^{-1}$ to 4000~cm$^{-1}$. Calculations are performed using the quantum scattering code from the \textrm{BIGOS} package developed in our group. The scattering S-matrix elements are obtained from boundary conditions imposed on the radial scattering function. Convergence of the calculated S-matrix elements is ensured by a proper choice of the integration range, propagator step, and the size of the rovibrational basis, see Appendix~\ref{app:convergence_uncertainty} for details.

Next, we calculate the generalized spectroscopic cross-sections, $\sigma^{q}_{\lambda}$~\citep{Monchick_1986,Schaefer_1992}, which describe how collisions perturb the shape of molecular resonance. Contrary to the state-to-state cross-sections, which give the rate coefficients (see, for instance,~\citet{Wan_2019}), the $\sigma^{q}_{\lambda}$ cross-sections are complex. For $\lambda=0$, real and imaginary parts of this cross-section correspond to the pressure broadening and shift cross-section, respectively. For $\lambda=1$, the complex cross-section describes the collisional perturbation of the translational motion and is crucial for the proper description of the Dicke effect. The index $q$ is the tensor rank of the spectral transition operator and equals 1 for electric dipole lines {considered here}.

We use the $\sigma^{1}_{0}$ and $\sigma^{1}_{1}$ cross-sections to calculate the six line-shape parameters relevant to collision-perturbed HD spectra, the collisional broadening and shift
\begin{equation}
    \gamma_{0} - i\delta_{0} = \frac{1}{2\pi c}\frac{1}{k_{B}T}\langle v_{r}\rangle \int_{0}^{\infty}  x e^{-x} \sigma^{1}_{0}(x)\mathrm{d}x,
\label{eq:gamma0anddelta0}
\end{equation}
the speed dependences of collisional broadening and shift
\begin{align}
\begin{split}
&\gamma_2-i\delta_2=\frac{1}{2\pi c}\frac{1}{k_B T}\frac{\left<v_r\right>\sqrt{M_a}}{2}e^{-y^2}\\ 
\times&\int\limits_{0}^{\infty}\left(2\bar{x}\cosh(2\bar{x}y)-\left(\frac{1}{y}+2 y\right)\sinh(2\bar{x}y)\right)\bar{x}^2e^{-\bar{x}^2}\sigma^{1}_{0}(\bar{x}\overline{v}_{p})d\bar{x},
\end{split}
\label{eq:gamma2anddelta2}
\end{align}
and the real and imaginary parts of the complex Dicke parameter 
\begin{align}
\begin{split}
        \tilde{\nu}_{opt}^r - & i \tilde{\nu}_{opt}^i = \frac{1}{2\pi c} \frac{\langle v_r \rangle M_a}{k_B T}  
        \int \limits_{0}^{\infty} x e^{-x} \left[\frac{2}{3}x\sigma_{1}^{1}(xk_B T) -\sigma_{0}^{1}(xk_B T)\right] dx,
\label{eq:nuopt}
\end{split}
\end{align}
where $v_r$ is the relative (absorber to perturber) speed of the colliding molecules, $\langle v_{r} \rangle$ is its mean value at temperature $T$, $\bar{v}_{p}$ is the most probable speed of the perturbed distribution, $M_{a} = \frac{m_{a}}{m_{a}+m_{p}}$, $x=\frac{E_{\mathrm{kin}}}{k_{B}T}$, $\bar{x}=\frac{2v_r}{\sqrt{\pi M_a}\left<v_r\right>}$, $y=\sqrt{\frac{m_p}{m_a}}$ and $m_{a}$ and $m_{p}$ are the masses of the active and perturbing molecules, respectively~(\cite{Wcislo_2021}). We estimate the uncertainty of the calculated line-shape parameters in Appendix~\ref{app:convergence_uncertainty}. {The six line-shape parameters define the modified Hartmann-Tran (mHT) profile (see Appendix~\ref{appendix:mHT}), which encapsulates {the relevant} beyond-Voigt effects.} To make the outcome of this work consistent with the HITRAN database~(\citet{Hitran_2020}), we provide temperature dependences of the calculated line-shape parameters within the double-power-law~(DPL) format~(\citet{Gamache_2018,Stolarczyk2020}):

\begin{align}
\begin{split}
        &\gamma_0(T) = g_0(T_{\rm{ref}}/T)^n + g_0' (T_{\rm{ref}}/T)^{n'}, \\
        &\delta_0(T) = d_0(T_{\rm{ref}}/T)^m + d_0' (T_{\rm{ref}}/T)^{m'}, \\
        &\gamma_2(T) = g_2(T_{\rm{ref}}/T)^j + g_2' (T_{\rm{ref}}/T)^{j'}, \\
        &\delta_2(T) = d_2(T_{\rm{ref}}/T)^k + d_2' (T_{\rm{ref}}/T)^{k'}, \\
        &\tilde{\nu}_{opt}^r(T) = r(T_{\rm{ref}}/T)^p + r'(T_{\rm{ref}}/T)^{p'}, \\
        &\tilde{\nu}_{opt}^i(T) = i(T_{\rm{ref}}/T)^q + i'(T_{\rm{ref}}/T)^{q'},
\end{split}
\label{eq:DPL}
\end{align}
where $T_{\rm{ref}} = 296$~K.  

\section{Results: line-shape parameters for the R(0), R(1) and R(2) lines in HD perturbed by a mixture of H$_2$ and He}
\label{sec:Results}
\begin{figure}[!ht]
    \centering
    \includegraphics[width=0.48\textwidth]{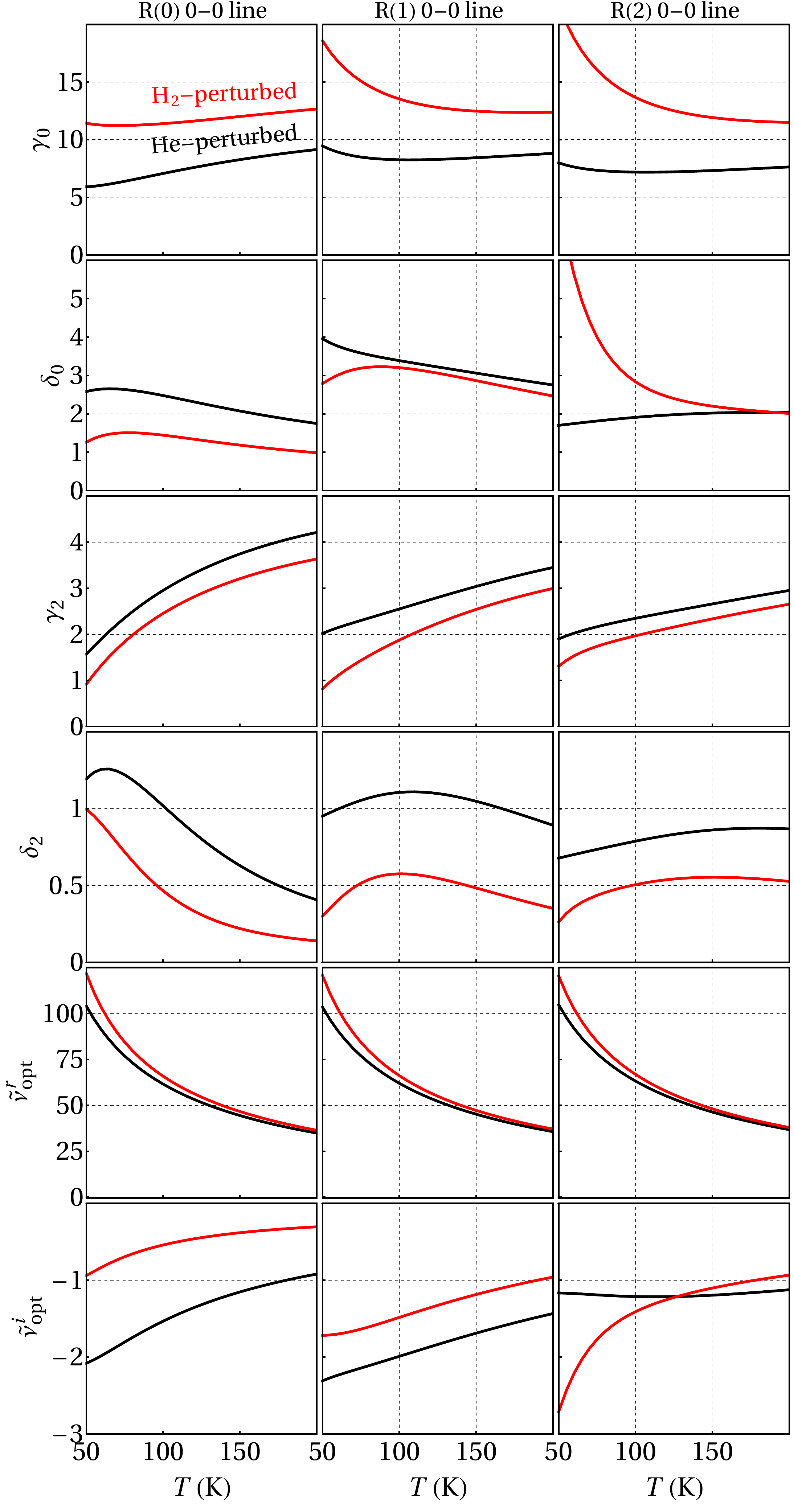}
    \caption{\textit{Ab initio} temperature dependences of the collisional line-shape parameters (in the units of 10$^{-3}$cm$^{-1}$atm$^{-1}$) of the first three electric dipole lines of HD perturbed by H$_2$ (red curves) and He (black curves).}
    \label{fig:perturbers}
\end{figure}

In Figure~\ref{fig:perturbers}, we show the main result of this work, i.e., all the six line-shape parameters for the R(0), R(1), and R(2) 0-0 lines in H$_2$-perturbed HD calculated as a function of temperature; Fig.~\ref{fig:perturbers} covers the temperature range relevant for the giant planets, 50 to 200~K (\citet{Lellouch_2001,Feuchtgruber_2013,Pierel_2017}), for our full temperature range, 20 to 1000~K, see the Supplementary Material. In Figure~\ref{fig:perturbers}, we also recall the corresponding He-perturbed data calculated with the same methodology at the same accuracy level (\cite{Stankiewicz_2021}). The difference between the two perturbers is not negligible, and for many cases the line-shape parameters differ by a factor of 2 or even more.

The data shown in Fig.~\ref{fig:perturbers} are given in Table~\ref{table:dataset} in a numerical form within the HITRAN DPL format, see Eqs.~(\ref{eq:DPL}). The accuracy of our \textit{ab initio} line-shape parameters is within 1\% of the magnitude of each parameter (see Appendix~\ref{app:convergence_uncertainty} for details). The DPL approximation of the temperature dependences introduces additional errors. For $\gamma_0(T)$ and $\tilde{\nu}_{opt}^r(T)$, the DPL error is negligible, for $\delta_0(T)$ the DPL error is at the 1~\% level, and for other line-shape parameters it can be even higher, but their impact on the final line profile is much smaller, see Appendix~\ref{appendix:DPL} for details. For applications that require the full accuracy of our \textit{ab initio} line-shape parameters, we provide the line-shape parameter values explicitly on a dense temperature grid in the {Supplementary Information}.

\begin{figure*}[t]
    \centering
    \includegraphics[width=\textwidth]{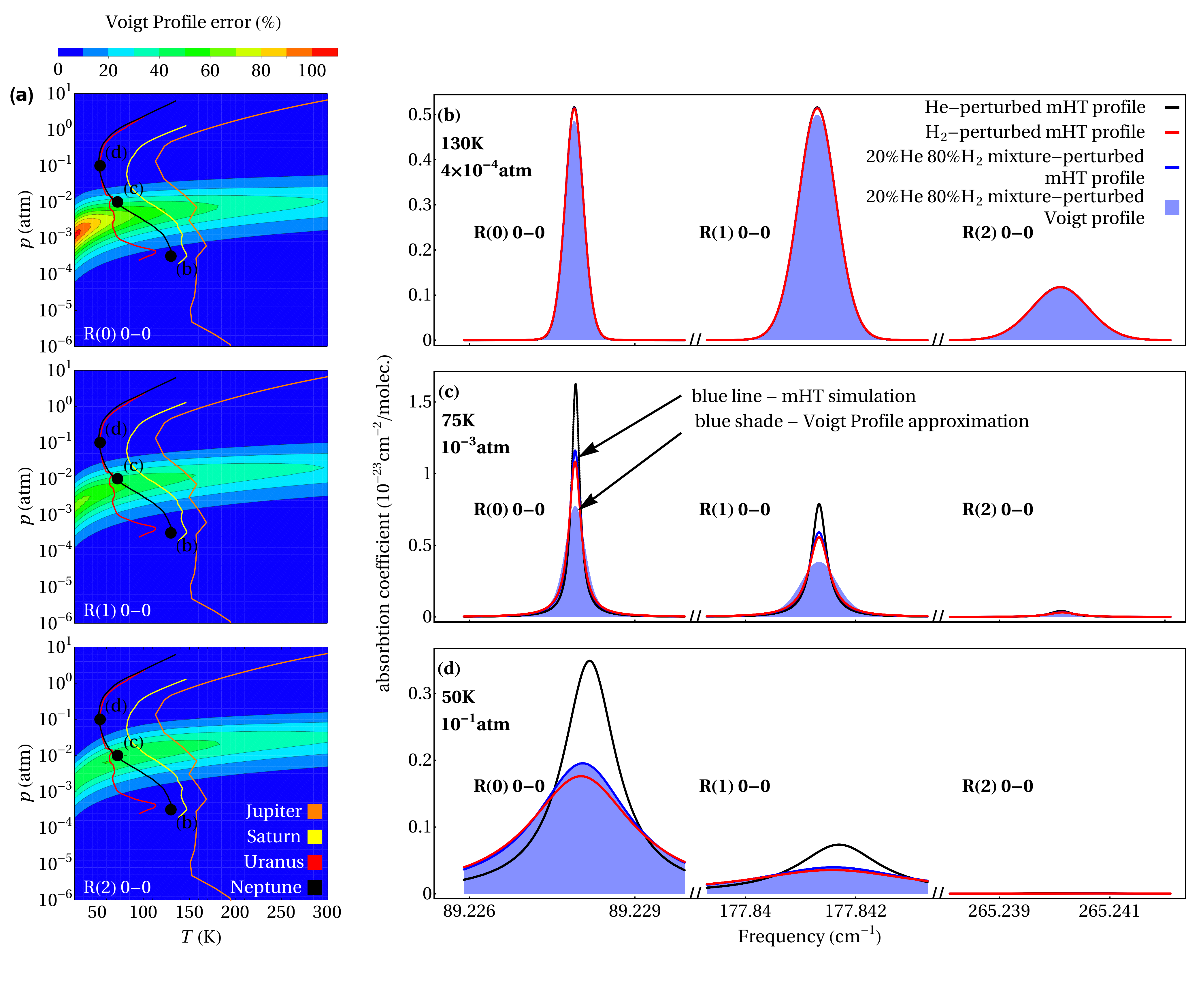}
    \caption{The role of the beyond-Voigt effects and bath mixture composition on collision-perturbed spectra of HD at conditions relevant for giant planet atmospheres. (a) Relative error of the Voigt-profile approximation as a function of pressure and temperature, shown as the relative difference between the Voigt and mHT profiles at profile maximum. The panels are arranged to correspond to the R(0), R(1), and R(2) lines, from top to bottom, respectively. (b), (c), (d) Simulations of the HD spectra (blue lines) at conditions relevant for the Neptune atmosphere (the perturbing bath consists of a mixture of 80\% H$_2$ and 20\% He). The spectra are generated with the mHT profile using HAPI based on the DPL temperature parametrization. As a reference, we show the same lines for the cases of pure H$_2$ and pure He perturbers, see the red and black lines, respectively. The blue shadows show the same simulations as the blue lines but generated with a simple Voigt profile. The three panels, (b), (c), (d), correspond to the three points, (b), (c), (d), shown in the temperature-pressure maps in panel (a) (the three selected points lay on the Neptune temperature-pressure line). The three cases illustrate three different line-shape regimes. The first one, (b), is the low-pressure case in which the lines are broadened mainly by the Doppler effect, and the pressure-induced collisional effects do not dominate the line shapes. The intermediate-pressure case, panel (c), illustrates the extreme non-Voigt regime (the differences between the blue curves and blue shadows reach almost a factor of 2), see also the green ridge in the maps in the bottom panel. The third case, panel (d), illustrates the high-pressure regime at which the HD lines are well described by the simple Voigt profile (the blue shadows almost overlap with the blue lines), but setting a proper composition of the perturber gas components plays an important role.}
    \label{fig:spectraSimulations}
\end{figure*}

The set of parameters in Table~\ref{table:dataset} contains all the information necessary to simulate the collision-perturbed shapes of the three HD lines at a high level of accuracy at any conditions relevant to the atmospheres of giant planets (pressure, temperature, and He/H$_2$ relative concentration). In Figure~\ref{fig:spectraSimulations}, we show an example of simulated spectra based on the data from Table~\ref{table:dataset}. It should be emphasized that at the conditions relevant to giant planets, the shapes of the HD lines may considerably deviate from the simple Voigt profile. In Figure~\ref{fig:spectraSimulations} (a), we show the difference between the Voigt profile and a more physical profile, which includes the relevant beyond-Voigt effects such as Dicke narrowing and speed dependence of broadening and shift (the modified Hartmann-Tran profile, see Appendix~\ref{appendix:mHT}). For the moderate pressures, the error introduced by the Voigt-profile approximation can reach almost 70\% (the orange, yellow, red, and black lines in Fig.~\ref{fig:spectraSimulations} (a) are the temperature-pressure profiles for Jupiter, Saturn, Uranus, and Neptune, respectively). We illustrate this with spectra simulations for the case of Neptune's atmosphere, see Figs.~\ref{fig:spectraSimulations} (b)-(d); the (b)-(d) panels corresponds to the (b)-(d) black dots in Fig.~\ref{fig:spectraSimulations} (a). The blue lines in Figs.~\ref{fig:spectraSimulations} (b)-(d) are the spectra simulated with the mHT profile and the blue shadows are the Voigt-profile approximations. The (b)-(d) cases illustrate three different regimes. The (b) case is the low-pressure regime with a small collisional contribution in which the line shape collapses to Gaussian, hence the beyond-Voigt effects are small. Case (d) is the opposite, the shape of a resonance is dominated by the collisional effects, but the simple Lorentzian broadening dominates over other collisional effects and again the beyond-Voigt effects are small. Case (c) illustrates the nontrivial situation in which the shape of resonance is dominated by the beyond-Voigt effects, see the green horizontal ridges in Fig.~\ref{fig:spectraSimulations} (a). The discrepancies between the beyond-Voigt line-shape model and the Voigt profile are also clearly seen as a difference between the blue lines and blue shadows in Fig.~\ref{fig:spectraSimulations} (c). In the context of the giant planet studies, it should be noted that the beyond-Voigt regions marked in Fig.~\ref{fig:spectraSimulations} (a) (the green horizontal ridges) coincide well with the maxima of the monochromatic contribution
functions for these three HD lines, see Fig.~1 in (\cite{Feuchtgruber_2013}) for the case of the atmospheres of Neptune and Uranus.

Figures~\ref{fig:spectraSimulations} (b)-(d) also illustrate the influence of atmosphere composition on the collision-induced shapes of the HD lines for the example of Neptune atmosphere. Black and red lines in Figs.~\ref{fig:spectraSimulations} (b)-(d) correspond to He- and H$_2$-perturbers, respectively. The gray lines correspond to the 20\% He and 80\% H$_2$ mixture that is relevant to Neptune atmosphere. The differences between black and red curves are negligible in the low-pressure regime, Figs.~\ref{fig:spectraSimulations} (b), since at these conditions the line shape is mainly determined by thermal Doppler broadening. At moderate- and high-pressure ranges, Figs.~\ref{fig:spectraSimulations} (c) and (d), the profiles differ at the peak by a factor of 2, hence including both perturbing species is important for spectra analysis of the atmospheres of giant planets, especially for the R(0) line, whose contribution function dominates at moderate and high pressures (\cite{Feuchtgruber_2013}). 

The data reported in this article, see Table~\ref{table:dataset}, account for three factors that are necessary for reaching sub-percent-level accuracy: 1) separate \textit{ab initio} data for both perturbers (that allows one to simulate perturbation by any H$_2$/He mixture), 2) accurate representation of temperature dependences, and 3) parametrization of the beyond-Voigt line-shape effects. In general, simulating the beyond-Voigt line-shape profiles is a complex task, see Appendix~\ref{appendix:mHT}. In this work, we used the HITRAN Application Programming Interface (HAPI) (\citet{Kochanov2016}) to generate the beyond-Voigt spectra shown in Fig.~\ref{fig:spectraSimulations} (based on the DPL parameters from Table~\ref{table:dataset}):
\begin{verbatim}
from hapi import *
db_begin ('hitran_data')
nu,coef = absorptionCoefficient_mHT(
    SourceTables='HD',
    Diluent={'He':0.2,'H2':0.8},
    WavenumberRange=[xmin,xmax],
    WavenumberStep=step,
    Environment={'p':press,'T':temp},
    HITRAN_units=True)
\end{verbatim}
The combination of the data reported in Table~\ref{table:dataset} and the Python-based HAPI constitutes a powerful tool that allows one to efficiently generate accurate HD spectra (based on advanced beyond-Voigt model, {the mHT profile}) for arbitrary temperature, pressure, and mixture composition.

\begin{figure}[!ht]
    \centering
    \includegraphics[width=0.98\linewidth]{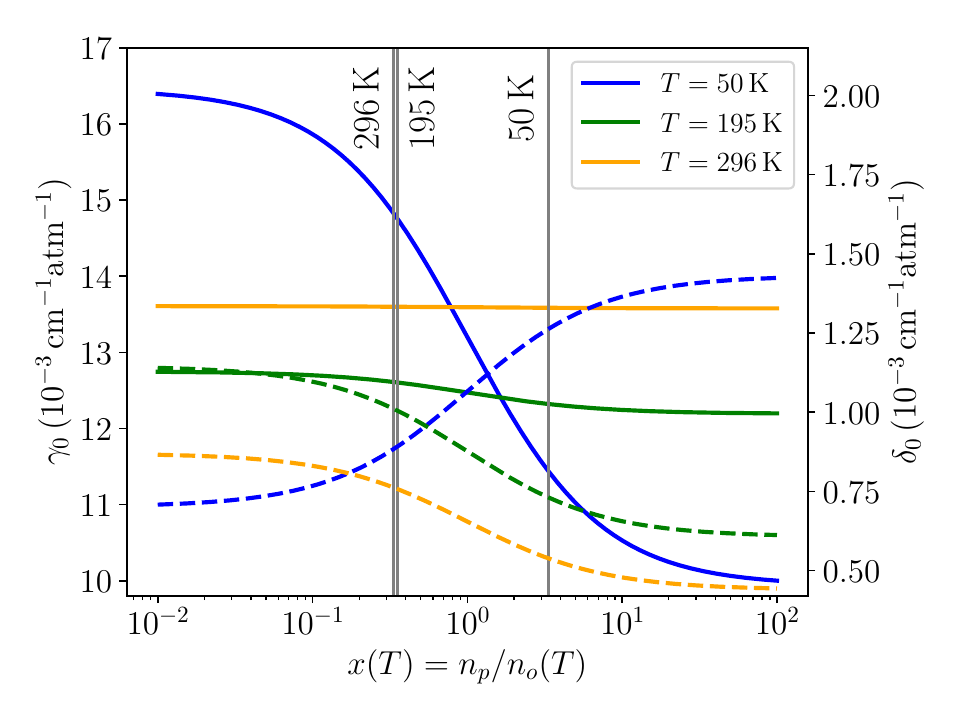}
    \caption{Spin isomer concentration ratio ($x=n_{p}/n_{o}$) dependence of pressure broadening, $\gamma_0$ (solid lines), and shift, $\delta_0$ (dashed lines), parameters for the H$_{2}$-perturbed R(0) line in HD, at different temperatures. The grey vertical lines correspond to the value of $x$ for normal H$_{2}$, $x=1/3$, and $x_{eq}(T)$, as determined by the Boltzmann distribution at $T=50$, $195$, and $296$~K.}
    \label{fig:spin:isomers}
\end{figure}

At low temperatures, relevant for studies of giant planet atmospheres and the chemistry and dynamics of the interstellar medium and protoplanetary discs, the spin isomer (\textit{para}/\textit{ortho}) concentration ratio of H$_{2}$ at thermal equilibrium (eq-H$_{2}$) deviates from 1:3 (the ratio of so-called normal H$_{2}$, n-H$_{2}$). Moreover, various processes, such as diffusion between atmospheric layers in gas giants, might result in the sub-equilibrium distribution of H$_{2}$. These non-trivial \textit{para}/\textit{ortho} distributions play a key role in atmospheric models that involve collisional induced absorption~(\citet{Karman_2019}) and spectral features originating from hydrogen dimers~(\citet{Fletcher_2018}), as well as in isotope chemistry of the interstellar medium, where \textit{para}/\textit{ortho} ratio controls the deuterium fractionation process~(\citet{Flower_2006, Nomura_2022}). In Figure~\ref{fig:spin:isomers}, we show the influence of the spin isomer concentration on the line-shape parameters. Spin isomer concentration has a large impact at low temperatures. All the line-shape parameters reported in this work are calculated for the thermal equilibrium spin isomer concentration. 

\FloatBarrier
  \clearpage
\begin{sidewaystable*}[!ht]
\centering
\caption{DPL parameterization of the temperature dependences of the line-shape parameters of HD perturbed by He and H$_2$. Coefficients 1 and 2 are in 10$^{-3}$cm$^{-1}$atm$^{-1}$. Exponents 1 and 2 are dimensionless.}
  \begin{minipage}[t]{0.48\linewidth} 
  \centering
		
   \begin{tabular}{ c r@{}l r@{}l r@{}l r@{}l }
\hline
\hline
\\[-1.5ex]
    &\multicolumn{8}{c}{He-perturbed HD}\\
    \cline{2-9}
    \multicolumn{9}{c}{ }\\
    &\multicolumn{8}{c}{R(0) 0-0 line}\\

    $\gamma_0(T)$ & 
    $g_0$&$=218.905$ & 
    $g'_0$&$=-209.398$&
    $n$&$=0.0929083$&
    $n'$&$=0.105086$\\[\mydistance]
    
    $\delta_0(T)$&
    $d_0$&$=76.1358$ & 
    $d'_0$&$=-74.6225$ & 
    $m$&$=-0.102987$ & 
    $m'$&$=-0.116673$\\[\mydistance]
    
     $\gamma_2(T)$&
    $g_2$&$=210.682$&
    $g'_2$&$=-206.11$&
    $j$&$=-0.862904$&
    $j'$&$=-0.876244$\\[\mydistance]
    
    $\delta_2(T)$ & 
    $d_2$&$=7.77569$ &
    $d'_2$&$=-7.45369$ &
    $k$&$=1.43509$ &
    $k'$&$=1.45106$\\[\mydistance]
    
     $\tilde{\nu}_{opt}^r(T)$ & 
     $r$&$=157.867$ & 
     $r'$&$=-132.815$ & 
     $p$&$=0.569778$ & 
     $p'$&$=0.512484$\\[\mydistance]

     $\tilde{\nu}_{opt}^i(T)$ & 
     $i$&$=-0.0148171$ & 
     $i'$&$=-0.732168$ & 
     $q$&$=0.589866$ & 
     $q'$&$=0.589866$\\
					
    \cline{2-9}
    \multicolumn{9}{c}{ }\\
    &\multicolumn{8}{c}{R(1) 0-0 line}\\

    $\gamma_0(T)$& 
     $g_0$&$=8.98598$ & 
     $g'_0$&$=0.07062$ & 
     $n$&$=-0.116463$ & 
     $n'$&$=1.83834$\\[\mydistance]

     $\delta_0(T)$ & 
     $d_0$&$=3.27647$ & 
     $d'_0$&$=-0.997538$ & 
     $m$&$=0.171067$ & 
     $m'$&$=-0.548127$\\[\mydistance]

    $\gamma_2(T)$ & 
    $g_2$&$=5.47783$ & 
    $g'_2$&$=-1.63047$ & 
    $j$&$=-0.540645$ & 
    $j'$&$=-1.19601$\\[\mydistance]

     $\delta_2(T)$ & 
     $d_2$&$=64.7968$ & 
     $d'_2$&$=-64.0672$ & 
     $k$&$=-0.314427$ & 
     $k'$&$=-0.325103$\\[\mydistance]

     $\tilde{\nu}_{opt}^r(T)$ & 
     $r$&$=39.7564$ & 
     $r'$&$=-13.9988$ & 
     $p$&$=0.652375$ & 
     $p'$&$=0.276761$\\[\mydistance]

     $\tilde{\nu}_{opt}^i(T)$ & 
     $i$&$=-0.0600157$ & 
     $i'$&$=-1.04666$ & 
     $q$&$=0.478987$ & 
     $q'$&$=0.478987$\\
      
    \cline{2-9}
    \multicolumn{9}{c}{ }\\
    &\multicolumn{8}{c}{R(2) 0-0 line}\\

    $\gamma_0(T)$ & 
    $g_0$&$=7.91606$ & 
    $g'_0$&$=0.106818$ & 
    $n$&$=-0.158512$ & 
    $n'$&$=1.59926$\\[\mydistance]

    $\delta_0(T)$ & 
     $d_0$&$=60.8755$ & 
     $d'_0$&$=-59.1362$ & 
     $m$&$=-0.560989$ & 
     $m'$&$=0.57118$\\[\mydistance]

    $\gamma_2(T)$ & 
    $g_2$&$=3.44668$ & 
    $g'_2$&$=-0.0918827$ & 
    $j$&$=-0.351514$ & 
    $j'$&$=-2.42064$\\[\mydistance]

    $\delta_2(T)$ & 
     $d_2$&$=48.9743$ & 
     $d'_2$&$=-48.2166$ & 
     $k$&$=-0.522538$ & 
     $k'$&$=-0.53522$\\[\mydistance]

    $\tilde{\nu}_{opt}^r(T)$ & 
     $r$&$=33.4337$ & 
     $r'$&$=-6.77194$ & 
     $p$&$=0.682741$ & 
     $p'$&$=0.0477307$\\[\mydistance]

     $\tilde{\nu}_{opt}^i(T)$ & 
     $i$&$=-0.160883$ & 
     $i'$&$=-0.705905$ & 
     $q$&$=0.278356$ & 
     $q'$&$=0.278356$\\
    \hline
    \end{tabular}	
  \end{minipage}
  \hfill
  \begin{minipage}[b]{0.48\linewidth} 
  \centering
   \begin{tabular}{r@{}l r@{}l r@{}l r@{}l }
\hline
\hline
\\[-1.5ex]
    \multicolumn{8}{c}{H$_2$-perturbed HD}\\
    \hline
    \multicolumn{8}{c}{ }\\
    \multicolumn{8}{c}{R(0) 0-0 line}\\
    $g_0$&$=13.1538$&
    $g'_0$&$=-0.0209275$&
    $n$&$=-0.102409$&
    $n'$&$=-4.47666$\\[\mydistance]
					 
    $d_0$&$=21.8124$&
    $d'_0$&$=-21.0255$&
    $m$&$=1.06916$&
    $m'$&$=1.08444$\\[\mydistance]

    $g_2$&$=215.242$ & 
    $g'_2$&$=-211.197$ & 
    $j$&$=-1.0114$ & 
    $j'$&$=-1.02533$\\[\mydistance]
					
     $d_2$&$=17.5611$ & 
     $d'_2$&$=-17.4505$ & 
     $k$&$=0.41528$ & 
     $k'$&$=0.405624$\\[\mydistance]
					
     $r$&$=25.0013$ & 
     $r'$&$=0.0005783$ & 
     $p$&$=0.889907$ & 
     $p'$&$=4.16595$\\[\mydistance]
					 
     $i$&$=-0.0049349$ & 
     $i'$&$=-0.257607$ & 
     $q$&$=0.64352$ & 
     $q'$&$=0.64352$\\
     
     \hline
    \multicolumn{8}{c}{ }\\
    \multicolumn{8}{c}{R(1) 0-0 line}\\

     $g_0$&$=11.5582$ & 
     $g'_0$&$=0.165258$ & 
     $n$&$=0.0781601$ & 
     $n'$&$=1.89474$\\[\mydistance]

    $d_0$&$=133.353$ & 
    $d'_0$&$=-131.231$ & 
    $m$&$=-0.244838$ & 
    $m'$&$=-0.257492$\\[\mydistance]

    $g_2$&$=131.359$ & 
    $g'_2$&$=-128.602$ & 
    $j$&$=0.484966$ & 
    $j'$&$=0.495567$\\[\mydistance]

    $d_2$&$=35.3179$ & 
    $d'_2$&$=-35.0081$ & 
    $k$&$=-0.330198$ & 
    $k'$&$=-0.33988$\\[\mydistance]

    $r$&$=12.373$ & 
    $r'$&$=12.3606$ & 
    $p$&$=1.02771$ & 
    $p'$&$=0.755534$\\[\mydistance]

    $i$&$=-0.307423$ & 
    $i'$&$=-0.307142$ & 
    $q$&$=0.656905$ & 
    $q'$&$=0.656905$\\
      
    \hline
    \multicolumn{8}{c}{ }\\
    \multicolumn{8}{c}{R(2) 0-0 line}\\

    $g_0$&$=5.05367$ & 
    $g'_0$&$=5.04862$ & 
    $n$&$=0.767269$ & 
    $n'$&$=-0.528791$\\[\mydistance]

    $d_0$&$=1.10007$ & 
    $d'_0$&$=0.346535$ & 
    $m$&$=0.0896926$ & 
    $m'$&$=1.57482$\\[\mydistance]

    $g_2$&$=4.9771$ & 
    $g'_2$&$=-1.86783$ & 
    $j$&$=-0.69627$ & 
    $j'$&$=-1.38558$\\[\mydistance]

    $d_2$&$=63.3239$ & 
    $d'_2$&$=-62.865$ & 
    $k$&$=-0.627633$ & 
    $k'$&$=-0.634526$\\[\mydistance]

    $r$&$=26.3022$ & 
    $r'$&$=0.0004301$ & 
    $p$&$=0.856557$ & 
    $p'$&$=4.29539$\\[\mydistance]

    $i$&$=-0.208711$ & 
    $i'$&$=-0.208511$ & 
    $q$&$=1.07071$ & 
    $q'$&$=1.07071$\\		
     \hline

			\end{tabular}
  \end{minipage}

		\label{table:dataset}	
  
\end{sidewaystable*}
\FloatBarrier

\section{Experimental validation}
\label{sec:Validation}

\begin{figure}[!ht]
    \centering
    \includegraphics[width=0.5\textwidth]{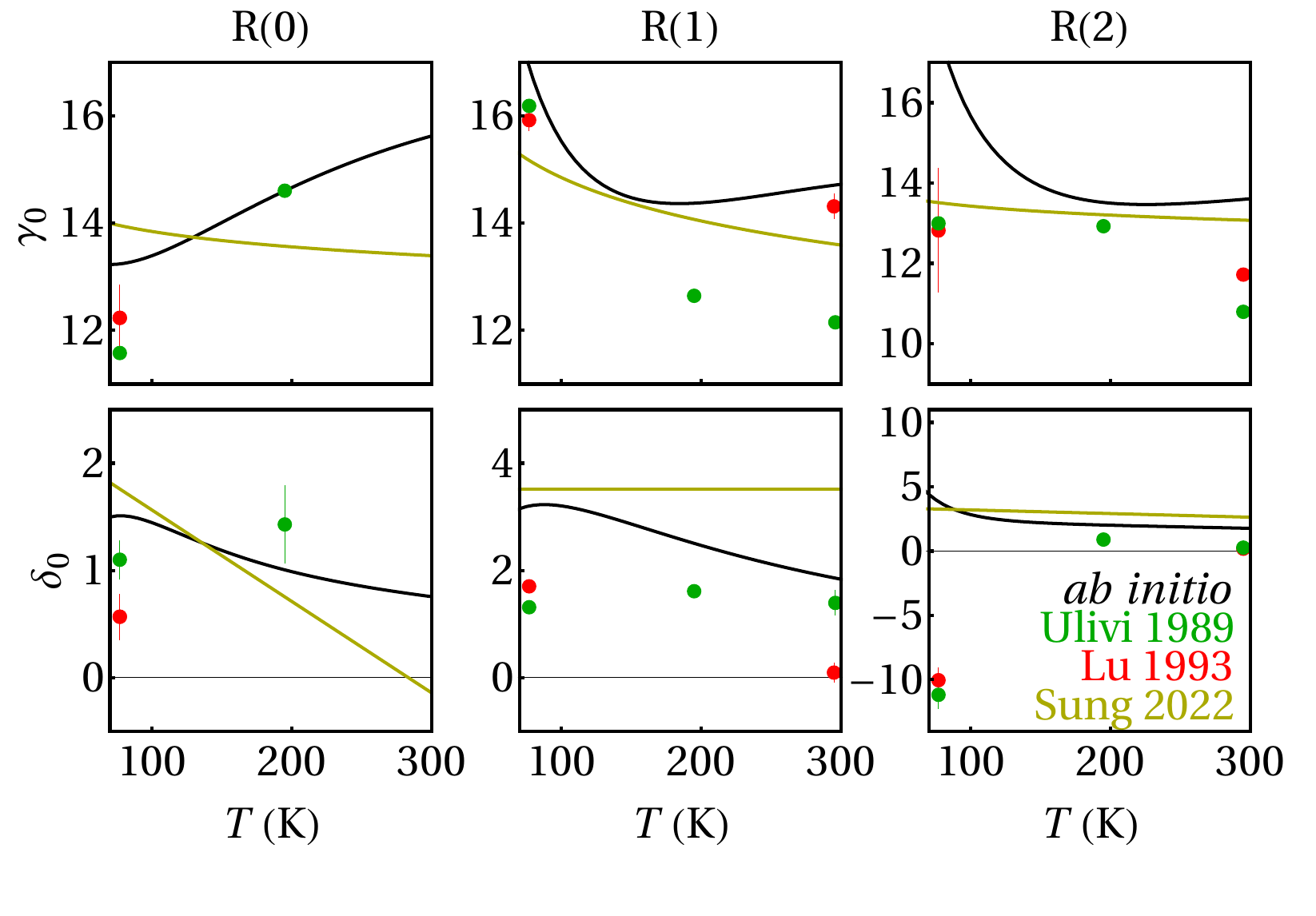}
    \caption{Comparison of the experimental and theoretical values of the pressure broadening and shift, $\gamma_0$ and $\delta_0$, parameters (in the units of 10$^{-3}$cm$^{-1}$atm$^{-1}$). Black curves correspond to the \textit{ab initio} calculations performed in this work, while green and red points report the experimental measurements from~(\cite{Ulivi_1989}) and~(\cite{Lu_1993}, respectively. The olive curves are the single-power-law (for $\gamma_0$) and linear (for $\delta_0$) temperature dependences retrieved from the measurements of~(\cite{Sung_2022}).}
    \label{fig:experimental}
\end{figure}

In Figure~\ref{fig:experimental}, we show a comparison between our \textit{ab initio} calculations (black lines) and the experimental data available in the literature. Fourier-transformed scans from the Michelson interferometer were used to obtain the high-pressure spectra reported in the works of \citet{Ulivi_1989} and \citet{Lu_1993}, see the green and red points, respectively; the spectra were collected in a temperature range from 77 to 296~K. Recently, the same lines were measured at low pressures (< 1 bar) with the Fourier transform spectrometer coupled to the Soleil-synchrotron far-infrared source (\cite{Sung_2022}) in a temperature range from 98 to 296~K, see the olive lines in Fig.~\ref{fig:experimental}. The discrepancy between these experimental data is by far too large to test our theoretical results at the one percent level. 

\begin{figure*}[!ht]
    \centering
    \includegraphics[width=\textwidth]{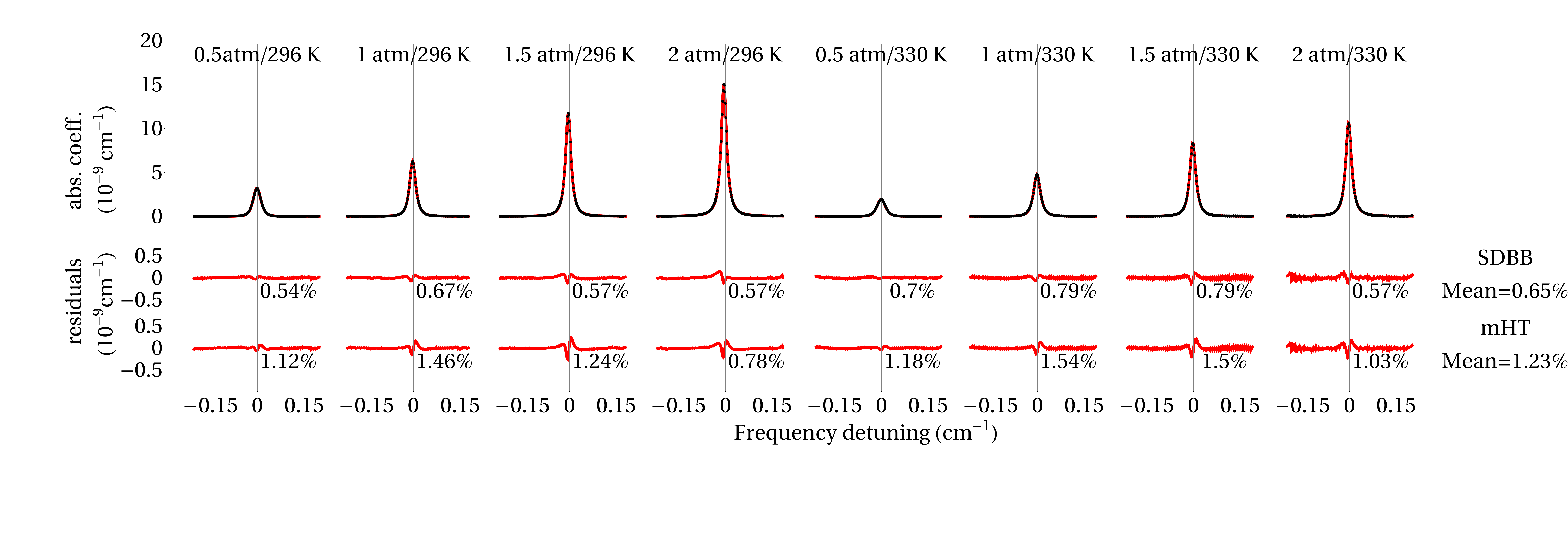}
    \caption{Direct validation of the \textit{ab initio} quantum-scattering calculations on the accurate experimental spectra of the S(2)~2-0 line of D$_2$ perturbed by collisions with H$_2$ molecules (see the text for details). The black dots are the experimental spectra and the red lines are the \textit{ab initio} profiles. Below each profile, we show the absolute residuals of two models: the speed-dependent billiard-ball (SDBB) profile and the modified Hartmann-Tran (mHT) profile. To quantify how well theory agrees with experiments, we report the relative (with respect to the profile peak value) root mean square errors (rRMSE) of the experiment-theory differences calculated within the $\pm$FWHM range around the line center, see the numbers (in percent) below the residuals. The mean rRMSE are also summarized for each of the models, see the numbers on the right side of the figure.}
    \label{fig:D2H2exp}
\end{figure*}

To validate our \textit{ab initio} calculations at the estimated accuracy level, we performed accurate measurements using a frequency-stabilized cavity ring-down spectrometer (FS-CRDS) linked to an optical frequency comb (OFC), referenced to a primary frequency standard~(\cite{Cygan_2016,Cygan_2019,Zaborowski_2020}). Our 73.5-cm-long ultrahigh finesse ($\mathcal{F}$ = 637~000) optical cavity operates in the frequency-agile, rapid scanning spectroscopy (FARS) mode~(\cite{Truong_2013,Cygan_2016,Cygan_2019}); see \citet{Zaborowski_2020} for details regarding the experimental setup. Since our spectrometer operates at 1.6~$\mu$m, we chose the S(2) 2-0 line in the H$_2$-perturbed D$_2$ (we repeated all the \textit{ab initio} calculations for this case). From the perspective of theoretical methodology, the H$_2$-perturbed D$_2$ and H$_2$-perturbed HD are equivalent and either can be used for validating the theoretical methodology (for both cases two distinguishable diatomic molecules are considered and the PES is the same except for the almost negligible DBOC term, see Appendix~\ref{appendix:PES}). We used a sample of 2\% D$_2$ and 98\% of H$_2$ mixture and collected the spectra at four pressures (0.5, 1, 1.5, and 2 atm) and two temperatures (296 and 330~K), see the black dots in Fig.~\ref{fig:D2H2exp}. The corresponding theoretical spectra are the red curves. The methodology for simulating the collision-perturbed shapes of molecular lines (based on the line-shape parameters calculated from Eqs.~(\ref{eq:gamma0anddelta0})-(\ref{eq:nuopt})) is described in {our previous works}~\citep{Wcislo_2018,Slowinski_2020,Slowinski_2022}. The two sets of residuals depicted in Fig.~\ref{fig:D2H2exp} show comparisons with two line-shape models, the speed-dependent billiard-ball profile (SDBB profile) and the mHT profile. The SDBB profile~\citep{Shapiro2002,Ciurylo2002} is the state-of-the-art approach that gives the most realistic description of the underlying collisional processes. As expected it gives the best agreement with experimental spectra (the mean residuals are 0.65\%, see Fig.~\ref{fig:D2H2exp}), but it is computationally very expensive (\cite{Wcislo_2013}). The mHT profile is slightly less accurate (the mean residuals are 1.23\%) but it is highly efficient from a computational perspective and, hence, well suited for practical spectroscopic applications. In conclusion, the \textit{ab initio} line-shape parameters reported in this work (Fig.~\ref{fig:perturbers} and Table~\ref{table:dataset}) lead to profiles that are in excellent agreement with accurate experimental spectra, and the theory-experiment comparison is limited by a choice of a line-shape model used to simulate the experimental spectra.

\section{Conclusion}
\label{sec:Conclusion}
We computed accurate collision-induced line-shape parameters for the three pure rotational HD lines (R(0), R(1), and R(2)), which are currently employed for the analysis of giant planets’ atmospheres. To this end, we investigated HD-H$_{2}$ collisions using coupled channel quantum scattering calculations on a new, highly accurate ab initio PES. Scattering S-matrices determined from these calculations allowed us to obtain the collisional width and shift, as well as their speed dependences and the complex Dicke parameter of H$_2$-perturbed HD lines. By integrating data from our previous work on the HD-He system~\citep{Stankiewicz_2020, Stankiewicz_2021}, we provide comprehensive results covering a wide range of thermodynamic conditions, including temperature, pressure, and H$_2$/He concentration, relevant to the atmospheres of giant planets. The theoretical methodology was validated through cavity ring-down spectroscopy, demonstrating sub-percent-level accuracy that surpasses the accuracy of previous theoretical and experimental studies of line-shape parameters in HD.

All the reported line-shape parameters and their temperature dependences are consistent with the HITRAN database format. Utilizing the HITRAN Application Programming Interface (HAPI), we demonstrated how our results can be applied to simulate HD spectra under various conditions pertinent to giant planets in the solar system. 

Until now, the analysis of observed collision-perturbed spectra in astrophysical studies has predominantly relied on the simple Voigt profile. We have introduced a methodology and provided a comprehensive dataset that enables the simulation of beyond-Voigt shapes for hydrogen deuteride in H$_{2}$/He atmospheres. Our work demonstrates that accounting for the speed dependence of collisional width and shift, along with the complex Dicke parameter, is crucial. These factors can alter the effective width and height of HD lines by up to a factor of 2, which in turn could significantly impact the HD abundance inferred from astrophysical observations.

\begin{acknowledgements}
H.J. and N.S. were supported by the National Science Centre in Poland through Project No. 2019/35/B/ST2/01118. H.J. was supported by the Foundation for Polish Science (FNP). K.S contribution is supported by budgetary funds within the Minister of Education and Science program "Perły Nauki", Project No. PN/01/0196/2022. D.L. was supported by the National Science Centre in Poland through Project No. 2020/39/B/ST2/00719. S.W. was supported by the National Science Centre in Poland through Project No. 2021/42/E/ST2/00152. P. J. was supported by the National Science Centre in Poland through Project No. 2017/25/B/ST4/01300. K.P. was supported by the U.S. National Science Foundation award CHE-1955328. K.Sz. was supported by the US NSF award CHE-23/3826. IEG's contribution was supported through NASA grant 80NSSC24K0080. P.W. was supported by the National Science Centre in Poland through Project No. 2022/46/E/ST2/00282. For the purpose of Open Access, the author has applied a CC-BY public copyright licence to any Author Accepted Manuscript (AAM) version arising from this submission. We gratefully acknowledge Polish high-performance computing infrastructure PLGrid (HPC Centers: ACK Cyfronet AGH, CI TASK) for providing computer facilities and support within
the computational grant, Grant No. PLG/2023/016409. Calculations have been carried out using resources provided by the Wroclaw Centre for Networking and Supercomputing (http://wcss.pl), Grant No. 546. The research is a part of the program of the National Laboratory
FAMO in Toruń, Poland.
\end{acknowledgements}

\bibliographystyle{aa}
\bibliography{bibliography.bib}

\begin{appendix}
\onecolumn

\section{Details of the PES calculations}
\label{appendix:PES}
{The 6D PES for the H$_{2}$-H$_{2}$ system was calculated at the following level of theory
\begin{align}
\begin{split}
E_{\rm int} =E_{\rm int}^{\rm HF} [{\rm 5}]
     + \delta E_{\rm int}^{\rm CCSD(T)} [{\rm Q5}] 
     + \delta E^{\rm T(Q)}_{\rm int}[{\rm Q}]+ \delta E^{\rm FCI}_{\rm int}[{\rm T}]
     + \delta E^{\rm DBOC}_{\rm int}[{\rm T}] .\\
\label{eq:Eint_0}
\end{split}
\end{align}
In all cases, the aug-cc-pV$X$Z basis sets \citep{Kendall:92} were employed with the cardinal number $X$ taking values 2 (D), 3 (T), 4 (Q), and 5. The consecutive terms are defined in the following way: $E_{\rm int}^{\rm HF} [{\rm 5}]$ is the interaction energy calculated at the Hartree-Fock (HF) level using the aug-cc-pV5Z basis set, $\delta E_{\rm int}^{\rm CCSD(T)} [{\rm Q5}]$ is the correlation contribution to the interaction energy calculated using the coupled-cluster method with up to perturbative triple excitations, CCSD(T), with the results extrapolated to CBS limits using the $1/X^3$  formula~\citep{Halkier:99} from the calculations in the aug-cc-pVQZ and aug-cc-pV5Z basis sets. The next contribution, $\delta E^{\rm T(Q)}_{\rm int}[{\rm Q}]= 
  E^{\rm CCSDT(Q)}_{\rm int}[{\rm Q}] - E^{\rm CCSD(T)}_{\rm int}[{\rm Q}]$, accounts for the electron correlation effects beyond CCSD(T) included in the CC method with up to perturbative quadruple excitations, CCSDT(Q), computed with the aug-cc-pVQZ basis set, whereas 
$\delta E^{\rm FCI}_{\rm int}[{\rm T}]= 
  E^{\rm FCI}_{\rm int}[{\rm T}] - E^{\rm CCSDT(Q)}_{\rm int}[{\rm T}]$  describes the electron correlation effects beyond CCSDT(Q), calculated with the aug-cc-pVTZ basis set. The diagonal Born-Oppen\-hei\-mer correction (DBOC)~\citep{Handy:86}, $\delta E^{\rm DBOC}_{\rm int}[{\rm T}]$, was calculated with the masses of $^{1}$H with the CCSD densities \citep{Valeev:03,Gauss:06} obtained with the aug-cc-pVTZ basis set. The $\delta E_{\rm int}^{\rm DBOC}$ term is the only one which depends on masses.   However, it is small compared to other terms and would be even smaller if calculated for HD--H$_2$ instead of H$_2$--H$_2$. Thus, the resulting surface can be applied to the former system with 
full confidence.}
  
  {The interaction energies were fitted by an analytic function that consisted of short- and long-range parts, with a smooth switching at $R$ values between 9 and 10 $a_{0}$, using the switching function from~\citet{Babin:13b}. The short-range part was taken in the form of a sum of products of exponentials ${\rm e}^{-\alpha r_{ab}}$, where $r_{ab}$ are atom-atom distances~\citep{Fernandez:99,Braams:09}. In contrast to most published work that uses the same $\alpha$ for all terms, we used four different optimized values. The form of the long-range part was taken from~\citet{Patkowski:08}, but the parameters $C_{n}^{l_{1}l_{2}l}$ were multiplied by linear combinations of symmetry-invariant polynomials of $r_1$, $r_2$, and of magnitudes
  of their differences. The linear coefficients are determined from the fit to the {\em ab initio} energies obtained at the same level of theory as for the short-range part. The PES is expected to be valid for $r_i \in [0.85, 2.25] \,a_{0}$. \citet{Zuo:21} recently published a 6D PES for the H$_2$ dimer obtained from the complete active space self-consistent field (CASSCF) calculations combined with the multi-reference configuration interaction (MRCI) calculations. According to~\citet{Zuo:21}, the potential is valid up to $r_i = 3.45 \,a_{0}$, beyond the upper limit of our PES. However, in the region of validity of both PESs, our surface should be more accurate due to the higher level of theory used.}

\section{Details of quantum scattering calculations}
\label{app:quantum_scattering}
The dependence of the 6D PES on the three Jacobi angles is separated from the radial and intramolecular distances by the expansion of the PES over the bispherical harmonics, $I_{l_1 l_2 l}(\theta_1, \theta_2, \phi)$
\begin{equation}
\label{eq:PESexpansion}
    V(R, r_{1}, r_{2}, \theta_1, \theta_2, \phi) = \sum_{l_1,l_2,l} A_{l_1 l_2 l}(R, r_{1}, r_{2}) I_{l_1 l_2 l}(\theta_1, \theta_2, \phi) ,
\end{equation}
where the bispherical harmonics are defined as
\begin{eqnarray} \label{eq:biharm}
    I_{l_1 l_2 l}(\theta_1, \theta_2, \phi = \phi_{1} - \phi_{2}) = \sqrt{\frac{2l+1}{4\pi}} \sum_{m} (l_1\,m \,l_2\,{-m}\,|\,l_1\,l_2\,l\,0)  Y_{l_1 m}(\theta_1, \phi_1) Y_{l_2 {-m}}(\theta_2, \phi_2) .
\end{eqnarray}
If the $i$-th ($i=1, 2$) molecule is homonuclear, the $l_{i}$ index in the expansion in Eq.~\eqref{eq:PESexpansion} takes only even values.
The $A_{l_1 l_2 l}(R, r_{1}, r_{2})$ expansion coefficients are obtained by integrating the product of the PES and the corresponding bispherical harmonic, over Jacobi angles (see, for instance, Eq.~(3) in~\citet{Zadrozny_2022} and the discussion therein). We employ a 19-point Gauss-Legendre quadrature to integrate over $\theta_{1}$ and $\theta_{2}$ and a 19-point Simpson's rule for the integral over $\phi$. The integration results in a tabular representation of the $A_{l_1 l_2 l}(R, r_{1}, r_{2})$ expansion coefficients, calculated for $R$ in the range of 2.5 to 200~$a_{0}$ with a step of 0.1~$a_{0}$, and for intramolecular distances ranging from 0.85 to 2.25~$a_{0}$ with a step of 0.1~$a_{0}$.

In this work, we use terms up to the $I_{4 4 8}(\theta_1, \theta_2, \phi) $ bispherical harmonic, which corresponds to the total number of 19 and 32 terms in the D$_{2}$-H$_{2}$ and HD-H$_{2}$ case, respectively. Such a number of terms represents an intermediate complexity of the problem -- the number of terms in the HD-H$_{2}$ case is larger by a factor of four in comparison to the HD-He case~(\citet{Stankiewicz_2020,Stankiewicz_2021}), but significantly smaller in comparison to more anisotropic PESs, studied in our previous works (85 for O$_{2}$-N$_{2}$~(\citet{Gancewski_2021}), and 205 for CO-N$_{2}$ and CO-O$_{2}$~(\citet{Jozwiak_2021,Zadrozny_2022})). The error introduced by the truncation of the PES expansion is discussed in Appendix~C.

The dependence of the expansion coefficients on $r_{1}$ and $r_{2}$ is reduced by averaging $A_{l_1 l_2 l}(R, r_{1}, r_{2})$ over rovibrational wave functions of isolated molecules, ($\chi_{\eta_{i}}(r_{i})$)
\begin{eqnarray}
\label{eq:PESaverage}
    &&A_{l_1 l_2 l, \eta_{1}, \eta_{1}^{'}, \eta_{2}, \eta_{2}^{'}} (R)=   \int \mathrm{d}r_{2} \chi_{\eta_{2}}(r_{2}) \Biggl( \int \mathrm{d}r_{1} \chi_{\eta_{1}}(r_{1}) A_{l_1 l_2 l} (R, r_{1}, r_{2}) \chi_{\eta_{1}'}(r_{1}) \Biggr) \chi_{\eta_{2}'}(r_{2}) ,
\end{eqnarray}
where $\eta_{i} = (\nu_{i}, j_{i})$ denotes the quantum numbers of a rovibrational state of the $i$-th molecule. The wave functions of H$_{2}$, HD, and D$_{2}$ are obtained by solving the nuclear Schr\"{o}dinger equation for isolated molecules using the potential energy curve of~\citet{Schwenke_1988}. We use the standard trapezoidal rule to perform the integration in Eq.~\eqref{eq:PESaverage}, and we obtain the $A_{l_1 l_2 l, \eta_{1}, \eta_{1}^{'}, \eta_{2}, \eta_{2}^{'}} (R)$ coupling terms for $R$ within the range 2.5 to 200~$a_{0}$, with a step size of 0.1~$a_{0}$.

The average in Eq.~\eqref{eq:PESaverage} provides a large number of possible coupling terms. In the HD-H$_{2}$ case, we consider pure rotational transitions (up to 1000~K), thus we neglect the terms that couple excited vibrational states ($\nu_{\rm{HD}}' \neq \nu_{\rm{HD}}$, $\nu_{\rm{H_{2}}}' \neq \nu_{\rm{H_{2}}}$). In the D$_{2}$-H$_{2}$ case, we observe that the terms that couple different vibrational levels are three orders of magnitude smaller than terms diagonal in $\nu$. Since we perform quantum scattering calculations in the $\nu_{\rm{D_{2}}} = 0$ and $\nu_{\rm{D_{2}}} = 2$ states separately (while maintaining $\nu_{\rm{H_{2}}} = 0$), we neglect radial coupling terms off-diagonal in vibrational quantum numbers. This approximation is additionally justified by the fact that the 2-0 S(2) line in H$_{2}$-perturbed deuterium is measured at room temperature, where the population of H$_{2}$ in vibrationally excited states is negligible. 

The dependence of the coupling terms in Eq.~\eqref{eq:PESaverage} on rotational quantum numbers (usually at the level of a few percent) is one of the key factors that affect theoretical predictions of the pressure shift of pure rotational lines in light molecules~(\citet{Shafer_1973,Dubernet_1999,Thibault_2016,Jozwiak_2018}). This is due to the fact that the line shift is sensitive to the difference in the scattering amplitude in the two rotational states which participate in an optical transition. Some authors neglect the $j$-dependence of the radial coupling terms (the centrifugal distortion of the PES) in scattering calculations for rovibrational transitions~(\citet{Green_1989,Thibault_2017}) and average the expansion coefficients for a given vibrational $\nu$ over the rovibrational wavefunction $\nu, j=0$. This approximation is invoked either to save computational resources or due to a lack of information about the dependence of the PES on the stretching coordinates but works well for Q($j$) lines. We have recently shown that taking into account centrifugal distortion of the PES is crucial for achieving a sub-percent agreement with the experimental spectra in He-perturbed vibrational lines in H$_{2}$~(\citet{Slowinski_2022}) and HD~(\citet{Stankiewicz_2020}). Thus, in both the HD-H$_{2}$ and D$_{2}$-H$_{2}$ cases, we include centrifugal distortion of the PES in scattering calculations. This leads to a large number of coupling terms (22~960 for $\nu=0$ state in HD, 16~359 in $\nu=0$ of D$_{2}$ and 17~157 for $\nu=2$), which is an order of magnitude more than in the case of HD-He (1~029). Similar to the HD-He case, this effect is crucial for achieving a sub-percent agreement with the cavity-enhanced spectra.

\section{Convergence of quantum scattering calculations and uncertainty budget for line-shape parameters}
\label{app:convergence_uncertainty}
\begin{table*}[!ht]
\caption{Summary of uncertainties associated with each convergence parameter in the quantum scattering calculations. The slashed values in the "Basis set size" line correspond to the case of collisions with \textit{para}-H$_{2}$ and \textit{ortho}-H$_{2}$, respectively. See the text for the details.}
\label{tab:uncertainties}
\begin{center}
\begin{tabular}{lcccccccc}
\hline
\multirow{2}{*}{Parameter} & \multirow{2}{*}{Value} & \multirow{2}{*}{Reference value} & \multicolumn{6}{c}{Maximum relative uncertainty (\%)} \\
& & & $\gamma_0$ & $\delta_0$ & $\gamma_2$ & $\delta_2$ & Re($\nu_{\text{opt}}$) & Im($\nu_{\text{opt}}$) \\
\hline
$R_{\rm{max}}$ & $100\,a_{0}$ &  $200\,a_{0}$ & 0.01 & 1 & 0.01 & 1 & 0.001 & 0.5 \\
$N_{\rm{steps}}$ & 200/100/50 & 500 & 0.4 & 1 & 0.5 & 3 & 0.1 & 5 \\
($l_{1}^{\rm{max}}, l_{2}^{\rm{max}}, l_{12}^{\rm{max}}$) in Eq.~\eqref{eq:PESexpansion} & (4,4,8) & (6,6,12) & 0.01 & 0.1 & 0.02 & 1 & 0.01 & 0.2 \\ 
Basis set size ($j_{\rm{HD}}^{\rm{max}}, j_{\rm{H_{2}}}^{\rm{max}}$) & see text & (6,6)/(7,7) & 0.01 & 0.2 & 0.01 & 0.2 & 0.01 & 0.2 \\
\hline 
Total & & & 0.4 & 1.4 & 0.5 & 3.5 & 0.1 & 5 \\ 
\hline
\end{tabular}
\end{center}
\end{table*}

In this appendix, we provide a detailed analysis of the convergence of generalized spectroscopic cross-sections with respect to specific parameters and how these parameters influence the uncertainty of the line-shape parameters. They include the range of propagation, the propagator step, the number of terms in the expansion of the PES~\eqref{eq:PESexpansion}, the size of the rotational basis set, and the number of partial waves. The uncertainties were estimated by calculating six line-shape parameters for the R(0) line at temperatures ranging from 20 to 1000 K. These values were then compared with those obtained from the generalized spectroscopic cross-sections calculated with the reference values of each parameter, which are significantly larger than the ones used in the final calculations. The stated uncertainties represent the maximum error observed within this temperature range. We assume that a similar level of accuracy is maintained for all other transitions considered in this paper. The results are summarized in Tab.~\ref{tab:uncertainties}.

The range of propagation is defined by the starting and ending points ($R_{\rm{min}}$ and $R_{\rm{max}}$). The smallest $R$ value that can be reliably calculated using quantum chemistry methods was employed as $R_{\rm{min}}$, which in this case (taking into account the coordinate transformation from the H$_{2}$-H$_{2}$ to HD-H$_{2}$ system) was $3\,a_{0}$. On the other hand, $R_{\rm{max}}$ should be large enough to apply boundary conditions on the scattering equations, i.e., in the range of $R$ where the PES becomes negligible compared to the centrifugal barrier. We tested the sensitivity of our results to $R_{\rm{max}}$ by performing calculations with $R_{\rm{max}}$ = 50~$a_{0}$, 75~$a_{0}$, 100~$a_{0}$, 150~$a_{0}$, and 200~$a_{0}$. After assessing the trade-off between computational cost and accuracy, we chose $R_{\rm{max}} = 100~a_{0}$ for the final calculations. Tab.~\ref{tab:uncertainties} provides uncertainties for six line-shape parameters impacted by this choice, estimated with respect to the reference calculations with $R_{\rm{max}} = 200~a_{0}$.

The step size of the propagation directly affects the precision and the computational cost of the quantum scattering calculations. We performed tests with a varying number of steps per half-de Broglie wavelength ($N_{\rm{steps}}$), including 10, 20, 30, 50, 100, 200, and 500. Based on these tests, we chose a step size of 50 for $E_{\rm{kin}} > 3$~cm$^{-1}$, 100 for $E_{\rm{kin}} \in (1.5, 3\rangle$~cm$^{-1}$, and 200 for $E_{\rm{kin}} \leq 1.5$~cm$^{-1}$. Uncertainties introduced by the choice of the number of steps, estimated with respect to the reference calculations with 500 steps per half-de Broglie wavelength, are gathered in Tab.~\ref{tab:uncertainties}.

We tested the convergence of the results with respect to the number of terms in the PES expansion (Eq.~\eqref{eq:PESexpansion}), comparing line-shape parameters for the R(0) line obtained from cross-sections calculated using a truncated expansion of the PES (with terms up to the $I_{4\,4\,8}(\theta_1, \theta_2, \phi) $ bispherical harmonic) and an expanded set of expansion coefficients describing higher anisotropies of the system (up to the $I_{6\,6\,12}(\theta_1, \theta_2, \phi)$ term). The results are gathered in Tab.~\ref{tab:uncertainties}.

The number of partial waves (or equivalently, blocks with given total angular momentum $J$) necessary to converge the scattering equations was determined based on a criterion of stability in the calculated cross-sections. We solved the coupled equations for an increasing number of $J$-blocks until four consecutive $J$-blocks contributed to the largest elastic and inelastic state-to-state cross-sections by less than $10^{-4}$~\AA$^{2}$. The convergence criterion ensured that the estimated error introduced by the number of partial waves was smaller than the smallest uncertainty attributable to the other parameters in our study. This implies that the uncertainty in the number of partial waves did not significantly contribute to the overall uncertainty in our results, thus we do not consider this factor in Tab.~\ref{tab:uncertainties}.

The size of the rotational basis set is a critical factor in quantum scattering calculations, and it was chosen with great care to ensure a consistent level of accuracy across different rotational states. For each calculation, we checked that the basis set included all energetically accessible (open) levels of the colliding pair, as well as a certain number of asymptotically energetically inaccessible (closed) levels. We gradually increased the size of the basis set until the calculated cross-sections did not show appreciable differences, identifying a fully converged basis set. We then determined the smallest basis set that ensured convergence to better than 1\% with respect to the fully converged basis. This was done for each initial state of the HD-H$_2$ system in a way that the estimated error for all transitions (R($j_{\rm{HD}}$), $j_{\rm{HD}}$=0, 1, 2), including those involving rotationally excited states, remained within the specified limit. 

The tests were conducted separately for collisions with \textit{para}-H$_{2}$ (which involves only even rotational quantum numbers) and \textit{ortho}-H$_{2}$ (which involves only odd rotational quantum numbers). In the case of \textit{para}-H$_{2}$, all rotational levels of HD and H$_{2}$ with $j \leq j^{\rm{max}} = 4$ were consistently included in the calculations. For specific calculations with H$_{2}$ initially in the $j_{\rm{H_{2}}}=4$ state, or HD initially in the $j_{\rm{HD}}=3$ state, the basis set was expanded to incorporate all rotational levels of HD and H$_{2}$ with $j \leq 6$. For \textit{ortho}-H$_{2}$, the basis set consistently included all rotational levels of HD and H$_{2}$ with $j \leq j^{\rm{max}}  =  5$. We extended the basis set to cover $j_{\rm{HD}}^{\rm{max}} = 6$ and $j_{\rm{H_{2}}}^{\rm{max}} = 5$ for cases where HD and H$_{2}$ were initially in the ($j_{\rm{HD}}, j_{\rm{H_{2}}}$) = (0,3), (1,3), or (2,1) states. The largest basis set, with $j_{\rm{HD}}^{\rm{max}} = j_{\rm{H_{2}}} = 7$, was employed in all calculations involving HD and H$_{2}$ in $(j_{\rm{HD}}, j_{\rm{H_{2}}})$ = (0,5), (1,5), (2,3), (2,5), (3,1), (3,3), and (3,5) states.

Assuming the uncertainties associated with each parameter as independent,  we estimated the maximum total uncertainty of each line-shape parameter using the root-sum-square method. The results are gathered in the last line of Tab.~\ref{tab:uncertainties}.

\section{Modified Hartmann-Tran (mHT) profile}
\label{appendix:mHT}
This appendix describes the modified Hartmann-Tran profile, which we used to simulate the spectra. We consider the mHT profile as the best compromise between the accurate but computationally-demanding speed-dependent billiard ball (SDBB) profile and the simple Voigt profile (VP). The mHT  profile can be expressed as a quotient of two quadratic speed-dependent Voigt (qSDV) profiles,
\begin{equation}
    \label{eq:mHT}
    \tilde{I}_{\mathrm{mHT}}(f)=\frac{\tilde{I}^{*}_{\mathrm{qSDV}}(f)}{1-(\nu_{opt}^r+i\nu_{opt}^i)\pi\tilde{I}^{*}_{\mathrm{qSDV}}(f)},
\end{equation}
which are directly linked to the spectral line-shape parameters from Eqs~\eqref{eq:gamma0anddelta0}-\eqref{eq:nuopt},
\begin{align}
    \label{eq:qSDV}
    \begin{split}
    \tilde{I}^{*}_{\mathrm{qSDV}}(f)=\frac{1}{\pi}\int  d^3\vec{v}f_m(\vec{v})
    \frac{1}{\Gamma_0+i\Delta_0+(\Gamma_2+i\Delta_2)(v^2/v^2_m-3/2)+\nu_{opt}^r+i\nu_{opt}^i-i(f-f_0-f_D v_z/v_m)}.
    \end{split}
\end{align}
The line-shape profile uses the parameters in the pressure-dependent form, i.e.,
\begin{align}
    \label{eq:parametersA}
    \begin{split}
    \Gamma_0 & =\gamma_0\cdot p,~ \Delta_0=\delta_0\cdot p\\
    \Gamma_2 & =\gamma_2\cdot p,~ \Delta_2=\delta_2\cdot p\\
    \nu_{opt}^r & =\tilde{\nu}_{opt}^r\cdot p,~ \nu_{opt}^i=\tilde{\nu}_{opt}^i\cdot p,
    \end{split}
\end{align}
$f_m(\vec{v})$ is the Maxwell-Boltzmann distribution of the active molecule velocity, $v_m$ is its most probable speed and $v_z$ is one of the three Cartesian components of the $\vec{v}$ vector. $f$, $f_{0}$, and $f_D$ are the frequency of light, the central frequency of the transition, and the Doppler frequency, respectively.
\par
The hard-collision model of the velocity-changing collisions, which is used in the mHT profile, suffices to describe the velocity-changing line-shape effects (such as the Dicke narrowing) in the majority of the molecular species. However, in the cases with a significant Dicke narrowing, such as molecular hydrogen transitions, the hard-collision model does not reproduce the line shapes at the required accuracy level. To overcome this problem, a simple analytical correction (the $\beta$ correction function) was introduced~(\cite{Wcislo_2016,Konefal2020}), which mimics the behavior of the billiard ball model and, hence, considerably improves the accuracy of the mHT profile for hydrogen, at negligible numerical cost. The correction is made by replacing the $\nu_{opt}^{r}$ with $\beta_\alpha(\chi) \nu_{opt}^{r}$, where $\alpha$ is the perturber-to-absorber mass ratio and $\chi=\nu_{opt}^{r}/\Gamma_D$ (where $\Gamma_D$ is the Doppler width, see \citet{Konefal2020} for details). It should be emphasized that the $\beta$ correction does not require any additional transition-specific parameters (it depends only on the perturber-to-absorber mass ratio  $\alpha$). The $\beta$ correction was applied every time the mHT profile was used in this work.

\section{DPL representation of the temperature dependences}
\label{appendix:DPL}

In this appendix, we discuss the details of the double power law (DPL) representation of the temperature dependences of the spectral line-shape parameters. The DPL function is used to convert the exact temperature dependence of the line-shape parameters into a simple, analytical expression, suitable for storing in spectroscopic databases~(\cite{Stolarczyk2020}). This conversion is done by fitting the DPL function to the actual \textit{ab initio} temperature dependence data. 
\par
Within this work, we calculate the \textit{ab initio} values at temperatures ranging from 50 to 1000~K. Due to its relevance to the atmospheres of giant planets, we chose to prioritize the 50-200~K temperature range and use the data only from this range to generate the DPL coefficients~(\cite{Stolarczyk2020}). Projection of the \textit{ab initio} data on the DPL representation is performed by fitting the DPL function in the selected temperature range. For the most faithful reconstruction of the temperature dependence, we perform several different fitting procedures (i.e. Newton, Quasi-Newton, Levenberg-Marquardt, global optimization, and gradient methods) and select the one that gives the best result (lowest rRMSE, see the next paragraph). This selection is done separately for each of the line-shape parameters and molecular transition. Mathematically, the two terms of the DPL functions are identical, thus to avoid swapping them, we followed the convention that the first base coefficients should always be greater than second base coefficients. Furthermore, to reduce the number of significant digits, we required that if the two base coefficients have opposite signs, their absolute values should differ at least by 1\textperthousand. 

\begin{table}[!ht]
    \centering
    \caption{Relative root mean square error (rRMSE) of the DPL representation. We fitted the DPL function (see Eqs.~\eqref{eq:DPL}) to the \textit{ab initio} data at the temperature range of 50-200~K. This table presents the root mean square differences between the actual \textit{ab initio} data and the DPL representation in this range (see Fig.~\ref{fig:DPL}), divided by the value of the line-shape parameter at 296~K. }
    \begin{tabular}{c c c c}
         \hline
         parameter & R(0) & R(1) & R(2) \\
         \hline
         $\gamma_0$& 0.04\% & 0.04\% & 0.03\% \\
         $\delta_0$& 0.48\% & 1.04\% & 0.29\% \\
         $\gamma_2$& 0.29\% & 0.20\% & 0.21\% \\
         $\delta_2$& 2.36\% & 4.40\% & 1.20\% \\
         $\tilde{\nu}_{\rm opt}^{r}$& 0.03\% & 0.03\% & 0.03\% \\
         $\tilde{\nu}^{i}_{\rm opt}$& 1.51\% & 1.39\% & 0.53\% \\
         \hline
    \end{tabular}
    \label{tab:DPL_rRMSE}
\end{table}

\begin{figure}[!ht]
    \centering
    \includegraphics[width=0.5\textwidth]{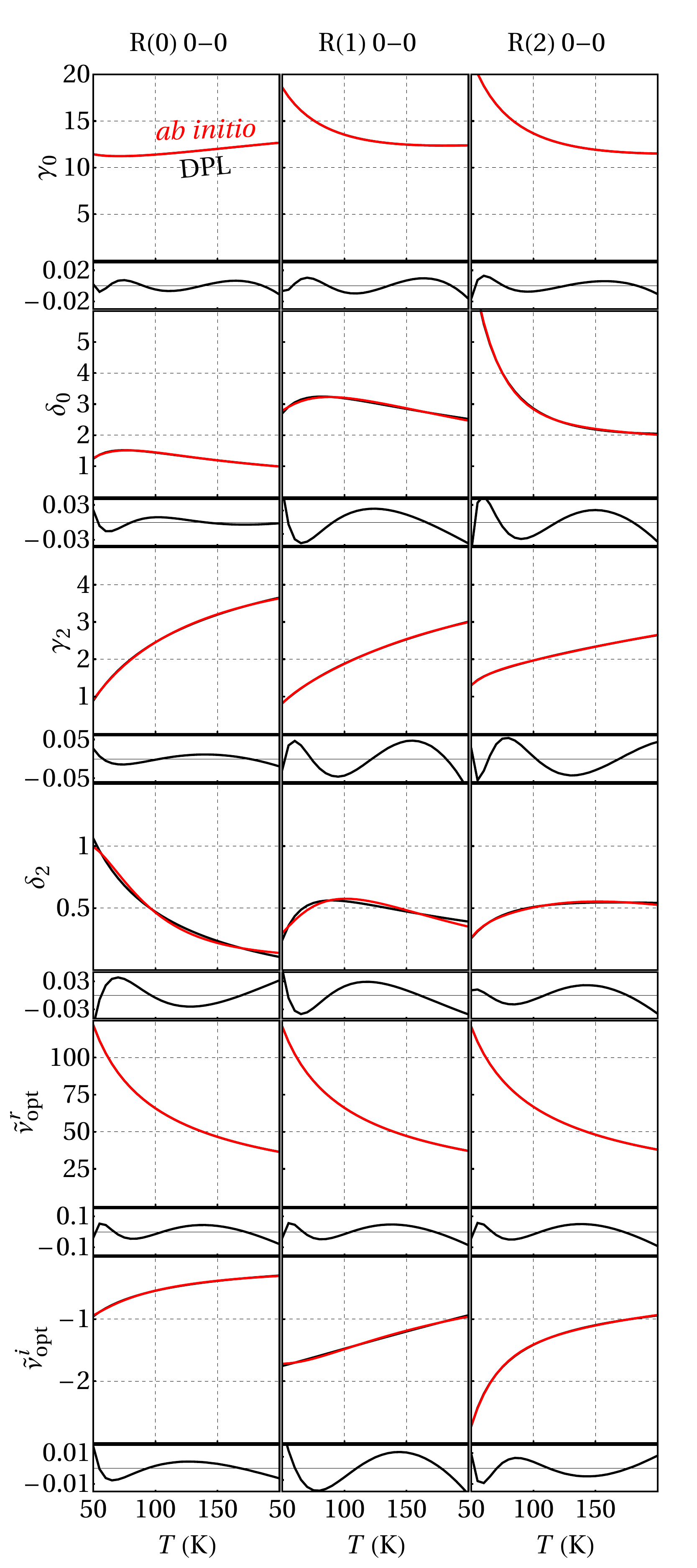}
    \caption{Temperature dependences of the six collisional line-shape parameters, $\gamma_0$, $\delta_0$, $\gamma_2$, $\delta_2$, $\tilde{\nu}_{\rm opt}^{r}$ and $\tilde{\nu}^{i}_{\rm opt}$ of the R(0)~0-0, R(1)~0-0 and R(2)~0-0 lines of HD perturbed by H$_2$. The red and black curves are the \textit{ab initio} results and DPL approximations, respectively. The small panels show the residuals from the DPL ﬁts. The vertical axes for all the panels (including residuals) are in 10$^{-3}$ cm$^{-1}$ atm$^{-1}$.}
    \label{fig:DPL}
\end{figure}

The DPL coefficients listed in Table~\ref{table:dataset} can be used to retrieve the temperature dependence of the line-shape parameters through Eqs.~\eqref{eq:DPL}. The efficiency of the DPL representation is depicted in Fig.~\ref{fig:DPL}. The red curves show the results of the \textit{ab initio} calculations, while the black curves (covered by the red ones in some cases) are the values reconstructed from the DPL coefficients from Table~\ref{table:dataset}. The corresponding residuals are presented under each of the plots. Table~\ref{tab:DPL_rRMSE} quantifies the accuracy of the DPL representation by presenting the values of the relative root mean square error (rRMSE) of the differences between the \textit{ab initio} data and the DPL fit. The values of the rRMSE are normalized with respect to the value of the corresponding line-shape parameter at 296~K. Even though the rRMSE values for some parameters are on the level of several percent, the overall error of the shape of the line is much smaller because the two parameters that impose the highest impact on the line width, $\gamma_0$ and $\tilde{\nu}_{\rm opt}^{r}$, are reproduced with high accuracy. For a detailed discussion on the propagation of errors from the parameters to the final shape of the line, see \citet{Slowinski_2022}.

\FloatBarrier
\FloatBarrier

\end{appendix}

\end{document}